\documentclass{ws-procs975x65}

\def\beq{\begin{equation}}
\def\eeq{\end{equation}}

\begin{document}

\title{Nonminimally coupled curvature-matter gravity models \\ and Solar System constraints}

\author{Riccardo March}

\address{Istituto per le Applicazioni del Calcolo, CNR, Via dei Taurini 19,\\
Roma, 00185, Italy\\
E-mail: r.march@iac.cnr.it}

\author{Orfeu Bertolami$^*$ and Jorge P\'aramos} 

\address{Departamento de F\'isica e Astronomia, Universidade do Porto, Rua do Campo Alegre 687,\\
Porto, 4169-007, Portugal\\
$^*$E-mail: orfeu.bertolami@fc.up.pt}

\author{Simone Dell'Agnello} 

\address{INFN, Laboratori Nazionali di Frascati (LNF), Via E. Fermi 40,\\
Frascati, 00044 Roma, Italy\\
E-mail: simone.dellagnello@lnf.infn.it}

\begin{abstract}
We discuss constraints to some nonminimally (NMC) coupled curvature-matter models of gravity by means of Solar System experiments. 

First we discuss a NMC gravity model which constitutes a natural extension of $1/R^n$ gravity to the nonminimally coupled case. 
Such a NMC gravity model is able to predict the observed accelerated expansion of the Universe. 
Differently from the $f(R)=1/R^n$ gravity case, which is not compatible with Solar System observations, 
it turns out that this NMC model is a viable theory of gravity.

Then we consider a further NMC gravity model which admits Minkowski spacetime as a background, 
and we derive the $1/c$ expansion of the metric.  
The nonrelativistic limit of the model is not Newtonian, but contains a Yukawa correction. 
We look for trajectories around a static, spherically symmetric body. Since in NMC gravity the energy-momentum tensor of matter is not conserved, 
then the trajectories deviate from geodesics.
We use the NMC gravity model to compute the perihelion precession of planets and
we constrain the parameters of the model from radar observations of Mercury.
\end{abstract}


\bodymatter


\section{Introduction}

We consider the possibility of constraining some nonminimally coupled (NMC) curvature-matter models of gravity \cite{BBHL} by means of Solar System experiments. 
The action functional involves two functions $f^1(R)$ and $f^2(R)$ of the Ricci curvature $R$. 
The function $f^1(R)$ is a nonlinear term which is analogous to $f(R)$ gravity, and the function $f^2(R)$ yields a NMC between the matter Lagrangian density 
and curvature. For other NMC gravity theories and their applications, see for instance \cite{PO1,PO2,PO3}.

 NMC gravity has been applied to several astrophysical and cosmological problems such as
dark matter \cite{dm1BP,dm2BFP}, cosmological perturbations \cite{pertBFP}, post-inflationary reheating \cite{reheating} or the current accelerated expansion of the Universe \cite{BFP}.

First we discuss the application of a perturbative method due to Chiba, Smith and Erickcek \cite{CSE}
to the NMC gravity model by Bertolami, Frazao and Paramos \cite{BFP}, which constitutes a natural extension of $1/R^n$ gravity to the non-minimally coupled case. 
Such a NMC gravity model is able to predict the observed accelerated expansion of the Universe. 
Differently from the $f(R)=R+1/R^n$ gravity case, which predicts the value $\gamma=1/2$ for the PPN parameter $\gamma$, so that the $f(R)$ model
is not compatible with Solar System observations, it turns out \cite{BMP} that the NMC gravity model
cannot be constrained, for specific choices of the functions $f^1(R)$ and $f^2(R)$, by the perturbative method considered by Chiba {\em et al.} \cite{CSE}, so that it remains, in this respect, a viable theory of gravity.

Then we consider a further NMC gravity model \cite{CPM,MPBDeA},
which admits Minkowski spacetime as a background, and we derive the $1/c$ expansion of the metric 
assuming the functions $f^1(R)$ and $f^2(R)$ analytic at $R=0$.  
The nonrelativistic limit of the model is not Newtonian, but contains a Yukawa correction. 
A parameterized post-Newton plus Yukawa (PPNY) approximation of the NMC model of gravity can be computed. 
We consider the metric around a static, spherically symmetric body and
we look for trajectories of a test body around the spherical body. Since in NMC gravity the energy-momentum tensor of matter is not conserved, 
then the trajectories deviate from geodesics. 
We use the NMC gravity model to compute the perihelion precession of planets. 
Eventually we constrain the parameters of the model from radar observations of Mercury,
including data from the NASA orbiter MESSENGER (MErcury Surface, Space ENvironment, GEochemistry and Ranging) spacecraft.


\section{The NMC gravity action functional}

The action functional of NMC gravity is given by \cite{BBHL}
\[
S = \int \left[\frac{1}{2}f^1(R) + [1 + f^2(R)] \mathcal{L}_m \right]\sqrt{-g}   d^4x,
\]
where $f^1(R),f^2(R)$ are functions of the spacetime curvature $R$, $g$ is the metric determinant,
$\mathcal{L}_m=-\rho c^2$ is the Lagrangian density of matter, and $\rho$ is mass density.

The function $f^2(R)$ yields a NMC between geometry and matter,
and the class of $f(R)$ gravity theories is recovered in the case $f^2(R)=0$.
General Relativity (GR) is recovered by taking:
\[
f^1(R) = 2\kappa(R-2\Lambda), \quad f^2(R) = 0, \quad \kappa = c^4/16\pi G,
\]
where $G$ is Newton's gravitational constant and $\Lambda$ is the Cosmological Constant.

The first variation of the action functional with respect to the metric yields the field equations
\begin{equation}\label{field-equations}
 \left(f^1_R + 2f^2_R \mathcal{L}_m \right) R_{\mu\nu} - \frac{1}{2} f^1 g_{\mu\nu} =
\nabla_{\mu\nu} \left(f^1_R + 2f^2_R \mathcal{L}_m \right)
+ \left(1 + f^2 \right) T_{\mu\nu},
\end{equation}
where $f^i_R = df^i\slash dR$ and
$\nabla_{\mu\nu}=\nabla_\mu \nabla_\nu -g_{\mu\nu}g^{\sigma\eta}\nabla_\sigma\nabla_\eta$.
Such equations will be solved by perturbative methods.


\section{A model for the accelerated expansion of the Universe}

We consider the NMC gravity model proposed by Bertolami, Frazao and Paramos \cite{BFP} to account for the observed accelerated expansion of the Universe:
\begin{equation}\label{NMC-model1}
f^1(R) = 2\kappa R, \qquad f^2(R) = \left( \frac{R}{R_n} \right)^{-n}, \quad n>0,
\end{equation}
where $n$ is an integer and $R_n$ is a constant. 
This NMC gravity model constitutes a natural extension to the non-minimally coupled case of the $1/R^n$ model proposed by Carroll {\em et al.} \cite{CDTT}
as an instance of $f(R)$ model.

Matter is described as a perfect fluid with negligible pressure \cite{BLP} with Lagrangian density $\mathcal{L}_m = -\rho c^2$.
We assume that the metric, which describes the spacetime around the Sun, is a perturbation of a flat Friedmann-Robertson-Walker (FRW) 
metric with scale factor $a(t)$:
\begin{equation}\label{metric}
ds^2 = -\left[1 + 2\Psi(r,t) \right] dt^2 + a^2(t)\left(\left[1 + 2\Phi(r,t)\right] dr^2
+ r^2 d\Omega^2 \right),
\end{equation}
where $|\Psi(r,t)| \ll 1$ and $|\Phi(r,t)| \ll 1$.
The NMC gravity model Eq. (\ref{NMC-model1}) yields a cosmological solution with a negative deceleration parameter $q <0$, and the scale factor $a(t)$ of the 
background metric follows the temporal evolution $a(t) = a_0 \left( t \slash t_0 \right)^{2(1+n)/3}$, where $t_0$ is the current age of the Universe \cite{BFP}.

In the perturbative approach developed by Chiba {\em et al.} \cite{CSE} for $f(R)$ gravity, the Ricci curvature of the perturbed spacetime is expressed as the sum
\[
R(r,t) = R_0(t) + R_1(r,t),
\]
where $R_0$ denotes the scalar curvature of the background FRW spacetime and $R_1$ is the perturbation due to
the Sun. The extension of the perturbative method of Chiba {\em et al.} \cite{CSE} to NMC gravity consists in the following steps \cite{BMP}.
We assume that functions $f^1(R)$ and $f^2(R)$ admit a Taylor expansion around $R=R_0$, and
we linearize the field equations (\ref{field-equations}) under two conditions: 
\begin{itemize}
\item [{\rm (i)}] terms nonlinear in $R_1$ can be neglected in the Taylor expansion of $f^1,f^2$;
\item[{\rm (ii)}] the following inequality
\begin{equation}\label{R1-cond}
\left\vert R_1(r,t) \right\vert \ll R_0(t),
\end{equation}
is satisfied both around and inside the Sun. 
\end{itemize}
We compute the functions $\Psi$ and $\Phi$ of the metric Eq. (\ref{metric}), then we find an expression
of the parameter $\gamma$ of the PPN (Parameterized Post-Newtonian) formalism \cite{Will}.
Eventually the validity of the condition (\ref{R1-cond}) is checked a posteriori.

The condition (\ref{R1-cond}) means that the curvature $R$ of the perturbed spacetime remains close to the cosmological value $R_0$ inside the Sun. 
In GR such a property of the curvature is not satisfied inside the Sun. However, for some $f(R)$ theories condition (\ref{R1-cond}) can be satisfied
and that leads to a violation of a constraint on PPN parameter $\gamma$ from Solar System tests of gravity.
For instance, the $1\slash R^n$ ($n>0$) gravity model \cite{CDTT} satisfies condition (\ref{R1-cond}) \cite{CSE,HMV}.

The perturbative solution of the field equations (\ref{field-equations}) yields the following expression
for the PPN parameter $\gamma=-\Phi(r)/\Psi(r)$ \cite{BMP}:
\[
\gamma = \frac{1}{2} \, \left[\frac{1 +  f^2_0 + 4 f^2_{R0}R_0 + 12 \square f^2_{R0}}
{1 +  f^2_0 +  f^2_{R0}R_0 + 3\square f^2_{R0}} \right],
\]
where $f^2_0=f^2(R_0)$ and $f^2_{R0} =d f^2/dR(R_0)$.
When $f^2(R)=0$ we find the known result $\gamma = 1\slash 2$ which holds for $f(R)$ gravity theories which
satisfy the condition $\left\vert R_1 \right\vert \ll R_0$ \cite{CSE}.
The $1\slash R^n$ ($n>0$) gravity theory \cite{CDTT}, where $f(R)$ is proportional to
$\left( R + {\rm constant}\slash R^n \right)$, is one of such theories that, consequently, have to be ruled out
by Cassini measurement \cite{Cassini}.

For the NMC gravity model (\ref{NMC-model1}), though $\left\vert R_1 \right\vert \ll R_0$ for $n \gg 1$, 
the solution for $R_1$ inside the Sun shows that non-linear terms in the Taylor expansion of $f^2(R)$ cannot be neglected \cite{BMP}:
\[
f^2(R) = f_0^2\bigg[ 1 - n \frac{ R_1}{R_0} + \frac{n(n+1)}{2} \left(\frac{R_1}{R_0}\right)^2  - \frac{1}{6}n(n+1)(n+2) \left(\frac{R_1}{R_0}\right)^3 \bigg] + O\left( \left( \frac{R_1}{R_0} \right)^4 \right).
\]
Hence assumption (i) is contradicted, implying the lack of validity of the perturbative regime. Eventually, by such a contradiction argument
the model (\ref{NMC-model1}) cannot be constrained by the extension to a NMC of the perturbative method by
Chiba {\em et al.} \cite{CSE}, so that the model (\ref{NMC-model1}) remains, in this respect, a viable theory of gravity \cite{BMP}.


\section{Planetary precession}

We now consider a NMC gravity model where the functions $f^1(R)$ and $f^2(R)$ are assumed analytic at $R=0$ \cite{MPBDeA}, so that they admit
the Taylor expansions:
\[
f^1(R) = 2\kappa \sum_{i=1}^\infty a_i R^i, \qquad a_1=1,
\qquad
f^2(R) = \sum_{j=1}^\infty q_j R^j.
\]
If $a_i=0$ for any $i>1$ and $q_j=0$ for any $j$, then the action of GR is recovered.

The model admits Minkowski spacetime as a background, and the $1/c$ expansion of the metric can be computed \cite{MPBDeA}, assuming
a general distribution of matter with mass density, pressure and velocity. 
The nonrelativistic limit of the model turns out to be non-Newtonian, but contains also a Yukawa correction. 
The coefficients $a_2,a_3,q_1,q_2$ are used to compute the metric at the order $O(1/c^4)$ for the $0-0$ component,
and are considered as parameters of the NMC gravity model.
A parameterized post-Newton plus Yukawa (PPNY) approximation of the NMC model of gravity can be computed \cite{MPBDeA}. 

Here we report the result \cite{MPBDeA} for the metric in vacuum around a static, spherically symmetric body (Sun) with uniform mass density ($g_{0i}=0$):
\begin{eqnarray}\label{metric-Sun}
g_{00} &=& -1 + 2 \frac{GM_S}{rc^2}\left( 1 + \alpha e^{-r/\lambda} \right)
+ \frac{2}{c^4}F(r), \nonumber\\
g_{ij} &=& \left[ 1 + 2 \frac{GM_S}{rc^2}\left( 1 - \alpha e^{-r/\lambda} \right) \right] \delta_{ij},
\end{eqnarray}
where $M_S$ is the mass of the spherical body, $F(r)$ is a radial potential, and $\lambda,\alpha$ are the range and strength of
the Yukawa potential which depend on the parameters of the NMC gravity model \cite{MPBDeA}:
\begin{equation}\label{Yukawa}
\lambda=\sqrt{6a_2}, \qquad
\alpha=\frac{1}{3}(1-\theta)+\frac{GM_S}{c^2R_S}\theta\left[ \theta\left(\frac{\mu}{2}-1\right)-
\frac{2}{3}\nu \right]\left(\frac{\lambda}{R_S}\right)^2 + \dots,
\end{equation}
where $R_S$ is the radius of the spherical body, $\theta,\mu,\nu$ are the following dimensionless ratios: $\theta=q_1/a_2$,
$\mu=a_3/a_2^2,\nu=q_2/a_2^2$, and dots $\dots$ denote smaller contributions \cite{MPBDeA}.
Formula (\ref{Yukawa}) has been obtained for $\lambda\gg R_S$.

Using the metric (\ref{metric-Sun}) the effect of NMC gravity on the orbit of a planet is
computed. In NMC gravity the energy-momentum tensor is not covariantly conserved\cite{BBHL}:
\[
\nabla_\mu T^{\mu\nu} = \frac{f^2_R }{ 1 + f_2} ( g^{\mu\nu} \mathcal{L}_m - T^{\mu\nu} ) \nabla_\mu R
\neq 0 \qquad\mbox{if }f^2(R)\neq 0,
\]
consequently, the trajectories deviate from geodesics:
\begin{equation}\label{geodesic}
\frac{d^2 x^\alpha}{ds^2} + \Gamma^\alpha_{\mu\nu} \frac{dx^\mu}{ds} \frac{dx^\nu}{ds} = \frac{f^2_R(R)}{ 1+f^2(R)} g^{\alpha\beta} R_{,\beta}.
\end{equation}
Moreover, geodesics are different from GR. The formula for perihelion precession of a planet has been computed \cite{MPBDeA} for
$\lambda\gg L$, where $L$ is the {\it semilatus rectum} of the unperturbed orbit. Here we report the leading term in the formula \cite{MPBDeA}:
\begin{eqnarray}\label{precession}
\delta\phi_P &=&\frac{6\pi GM_S}{Lc^2}+(1-\theta)^2\frac{\pi}{3}\left(\frac{L}{\lambda}\right)^2 e^{-L/\lambda} \nonumber\\
&+& (1-\theta)\frac{\pi GM_S}{3Lc^2}\theta\left[3\theta\left(\frac{\mu}{2}-1\right)-2\nu\right]
\left(1-\frac{L}{\lambda}\right)\left(\frac{L}{R_S}\right)^3+\dots,
\end{eqnarray}
where the terms in the first row are the GR precession and the nonrelativistic Yukawa precession, respectively, and
the term in the second row is the leading contribution from the NMC relativistic correction.
Dots $\dots$ denote smaller contributions \cite{MPBDeA}.
Eq. (\ref{precession}) reduces to the GR expression if $\theta=1$.

Using Eq. (\ref{precession}), bounds on PPN parameters from the Cassini experiment \cite{Cassini} and
fits to planetary data, including data from Messenger spacecraft \cite{Messenger} orbiting around Mercury,
it follows that the additional perihelion precession due to NMC deviations from GR, in the case of Mercury orbit, is bounded by \cite{MPBDeA}
\[
 - 5.87537 \times 10^{-4} < \delta \phi_P - 42.98'' < 2.96635 \times 10^{-3}.
\]
These inequalities define an admissible region in the four-dimensional parameter space with dimensionless coordinates
$\theta,\mu,\nu,R_S/\lambda$.
Exclusion plots obtained by slicing the admissible region with two-dimensional planes can be drawn \cite{MPBDeA}.

The admissible region in three-dimensional parameter subspace with coordinates $(\theta,\mu,\nu)$,
for $0<|1-\theta|\ll 1$ and a given $\lambda\gg L$, can be approximated by
the region enclosed within the degenerate quadric surfaces
\[
\nu=\frac{3}{4}\mu - \frac{3}{2} -9\left(\frac{R_S}{L}\right)^3 \frac{\varepsilon_i}{\left(1-L/\lambda\right)(1-\theta)}, \qquad i=1,2,
\]
where
\[
\varepsilon_1\,\frac{6\pi GM_S}{Lc^2} = - 5.87537 \times 10^{-4}, \qquad \varepsilon_2\,\frac{6\pi GM_S}{Lc^2} = 2.96635 \times 10^{-3}.
\]
The intersection of the three-dimensional admissible subregion with a plane $\theta=constant$, with $0<|1-\theta|\ll 1$,
is a strip enclosed between two lines in the $(\mu,\nu)$ plane. The intersections with the planes $\mu=constant$ and $\nu=constant$ are regions enclosed by pairs
of hyperbolae\cite{MPBDeA}.

Eventually, the BepiColombo mission to Mercury should allow for a reduction on the above bounds by approximately one order of magnitude \cite{BepiColombo}.

\section*{Acknowledgments}

The work of R.M. and S.DA is, respectively, partially and fully supported by INFN
(Istituto Nazionale di Fisica Nucleare, Italy), as part of the MoonLIGHT-2 experiment in the framework of the research 
activities of the Commissione Scientifica Nazionale n. 2 (CSN2).

%

\end{document}